\documentclass[journal]{IEEEtran}

\usepackage{amsmath,amssymb,graphicx,booktabs}
\usepackage{pgfplots}
\pgfplotsset{compat=1.18}
\usepgfplotslibrary{groupplots}
\usepackage[inline]{enumitem}
\usepackage{algorithm}
\usepackage{algpseudocode}
\usepackage{placeins} 
\usepackage[switch]{lineno}
\usepackage[normalem]{ulem}

\usepackage{subcaption}

\title{Study of Mixed-Integer Optimization Based on Graph-Based Decomposition for Cell-Free Networks}

\author{Julio Cesar Cardoso Tesolin and Rodrigo C.~de Lamare \vspace{-1.5em}}

\begin{document}
\maketitle

\begin{abstract}
This letter develops a radio access network (RAN) framework for
mixed discrete–continuous optimization problems that arise in user-centric cell-free massive multiple-antenna networks. The novel framework exploits the structural decomposition between discrete clustering decisions and continuous resource allocation variables by modeling the space of feasible serving states as a graph with Hamming-topology neighborhoods. A serving-state graph abstraction is introduced to enable topology-aware search-and-evaluate optimization procedures
and a graph-based search-and-evaluate (GBSE) algorithm is devised along with their complexity analysis. Energy efficiency maximization at the RAN level is presented as an application of considered alongside the proposed framework and GBSE algorithm. Numerical results show that minimal Hamming neighborhoods offer an attractive trade-off between scalability and exploration capability in graph-based optimization and GBSE outperforms existing techniques.
\end{abstract}


\section{Introduction}
\IEEEPARstart{F}{uture} mobile networks are expected to rely on clustered cell-free (CF) massive multiple-input multiple-output (mMIMO) architectures \cite{2021:ozlem:foundations,mmimo,wence} to enable distributed cooperation among access points (AP) while mitigating fronthaul constraints by assigning each user equipment (UE) to a subset of geographically relevant APs. In both user-centric and AP-centric formulations, the clustering problem plays a central role in balancing performance, scalability, and signaling overhead. To address this challenge, a wide range of optimization-based approaches have been proposed, adopting diverse network performance indicators as objective functions while jointly considering discrete clustering decisions and resource allocation variables. Regardless of the assumptions on objective functions or constraints, many of these formulations naturally fall within the class of mixed discrete–continuous optimization problems.

Several works have investigated clustering strategies using the channel norm or aimed at improving specific key performance indicators such as information rates. User-centric (UC) clustering approaches have been proposed to reduce signaling overhead and computational complexity \cite{ucentric,2022:mashdour:fairness,2022:mashdour:enhancedgreedy}, while other studies have focused on enhancing energy efficiency under UC formulations \cite{2023:ozlem:oran,2022:tan:hierarchical,2023:mendoza:drm,rra}. Despite these advances, the cluster selection problem remains inherently combinatorial, and existing solutions often rely on solver-specific techniques or heuristic rules without explicitly exploiting the structure of the discrete serving-state space. As a result, scalability remains a key limitation, particularly in large-scale CF deployments.

This letter proposes a solver-agnostic framework for UC clustering in CF-mMIMO networks. It is built upon: 
\begin{enumerate*}[label=(\roman*)]
    \item modeling the discrete space of radio access network (RAN) serving states as a structured Hamming-topology graph; 
    \item decomposing the resulting problem into a discrete outer-layer search and a inner-layer evaluation and 
    \item enabling a family of graph-based search algorithms that navigate the serving-state graph using local neighborhood information.
\end{enumerate*} 
Then, a graph-based search-and-evaluate (GBSE) algorithm is devised to solve problems using the proposed framework \cite{mio}. An energy-efficiency optimization at the RAN level is presented as to demonstrate the applicability of the framework and the GBSE algorithm.

{\it Notation}: Scalars are denoted by $a$, $A$; column vectors by $\mathbf{a} \in \mathbb{C}^n$; and matrices by $\mathbf{A} \in \mathbb{C}^{m \times n}$. $\mathcal{A}$ denotes a set and $\mathtt{A}$ denotes a graph. $(\cdot)^T$ is the transpose, $\mathbb{E}\{\cdot\}$ is the statistical expectation, and $\#(\cdot)$ is the set cardinality. $\mathcal{O}(\cdot)$ represents the worst-case asymptotic complexity.

\section{System Model}

We consider a UC CF-mMIMO network operating under time-division duplexing (TDD) and orthogonal frequency-division multiplexing (OFDM) \cite{2021:ozlem:foundations,c&didd,iddllr,iddocl,ccaps}. The network consists of $L$ access points (APs) connected to a central processing unit (CPU) with unlimited fronthaul bandwidth. Each AP is equipped with $N$ antennas, which cooperatively serve $K$ single-antenna UEs. Due to the envisioned ultra-dense characteristics, we assume that the total number of antennas in the network  significantly exceeds the total number of UEs, i.e., $NL \gg K$ \cite{2021:ozlem:foundations}.

\subsection{RAN As A Graph}\label{sec:rangraph}

Early CF-mMIMO studies relied on full cooperation, where all APs served all UEs, but this is impractical at scale and unnecessary when only a few APs provide meaningful channel gain. UC clustering restricts service to APs with sufficiently strong links, retaining near-optimal performance while reducing fronthaul load and per-AP processing. This flexible association is modeled using the dynamic cooperation clustering framework proposed in \cite{2021:ozlem:foundations}.

Let $\mathcal{D}_k \subseteq \mathcal{L}$ denote the non-empty set of APs $l\in\mathcal{L}=\{1,\dots,L\}$ eligible to serve UE $k\in\mathcal{K}=\{1,\dots,K\}$. These candidates may be selected based on physical-layer constraints (e.g., channel gain, pilot contamination), network-layer limitations (e.g., hardware constraints), or whether an AP $l$ positively contributes to a performance metric when serving UE $k$. 
This selection is captured by the binary decision variable $c_{k,l} \in \{0, 1\}$, collectively represented by the binary matrix $\mathbf{C} = [c_{k,l}] \in \{0, 1\}^{K \times L}$ where $\sum_{l=1}^L c_{k,l} \ge 1$.
To this end, we define $\mathcal{S}_k$ as the set of all subsets of $\mathcal{D}_k$ representing all possible AP combinations able to serve UE $k$, ranging from the traditional single-AP service to full cell-free cooperation. Considering $D_k=\#(\mathcal{D}_k)$, the cardinality of $\mathcal{S}_k$ is given by 
\begin{equation}
{S}_k =\#(\mathcal{S}_k) = \sum_{i=1}^{{D}_k} \binom{{D}_k}{i} = 2^{D_k} - 1.
\label{eq:clustercount}
\end{equation}

Let $\mathtt{B} = (\mathcal{K}, \mathcal{L}, \mathcal{E})$ denote a simple bipartite graph to model the serving association between UEs and APs, having $\mathbf{C}$ as its adjacency matrix \cite{mio,ldpc,memd}. The edge set $\mathcal{E} \subseteq \mathcal{K} \times \mathcal{L}$ satisfies the bipartite property $(k, l) \in \mathcal{E} \iff$ AP $l$ serves UE $k$.  For a given UE $k$, its local edge set is defined as $\mathcal{E}_k = \{(k, l) : l \in \mathcal{S}_{k,i}\}$ for some $\mathcal{S}_{k,i} \in \mathcal{S}_k, i \in \{1, \dots,S_k\}$. Thus, each element $\mathcal{S}_{k,i}$ induces a partial bipartite subgraph $\mathtt{B}_{k,i} = (k, \mathcal{S}_{k,i}, \mathcal{E}_k)$, connecting a UE $k$ to the APs in $\mathcal{S}_{k,i}$. The set of all possible bipartite graphs over the K UEs is obtained by selecting one serving set $\mathcal{S}_{k,i}$ for each UE $k$, given by $\mathcal{U} = \{\mathtt{B} = (\mathcal{K}, \mathcal{L}, \mathcal{E}) \mid \mathcal{E} = \bigcup_{k=1}^K (k, l) : l \in \mathcal{S}_{k,i},\text{ } \mathcal{S}_{k,i} \in \mathcal{S}_k,\text{ } i \in \{1, \dots, \mathcal{S}_{}k\}\}$. Thus, the total number of possible bipartite graphs or \textbf{RAN serving states} is described by
\begin{equation}
U=\#(\mathcal{U}) = \prod_{k=1}^{K}S_k= \prod_{k=1}^{K} \left(2^{{D}_k} - 1\right).
\label{eq:gclustercount}
\end{equation}
The envisioned increase in both the number of UEs and APs leads to a rapid escalation of $U$, as indicated by \eqref{eq:gclustercount}, having a lower and an upper bound given by $1 \leq U \leq (2^{L} - 1)^K$.
The lower bound represents the single-AP service ($D_k= 1, \forall k$), while the upper bound represents the full cell-free operation ($D_k= L, \forall k$). Thus, investigating each possible RAN serving state may become computationally prohibitive. This combinatorial nature underscores the need for a proper representation of $\mathcal{U}$ along with scalable searching algorithms that can efficiently approximate near-optimal configurations.

The exploration of the RAN serving states $\mathtt{B}$ in $\mathcal{U}$ to assess 
a given objective function 
calls again for the use of graph modeling. Hence, we denote $\mathtt{R}=(\mathcal{U},\mathcal{Q},\phi)$ an undirected graph, where $\mathtt{B} \in \mathcal{U}$ is a vertex, $(\mathtt{B},\mathtt{B}^{\prime}) \in \mathcal{Q}$ is an edge connecting two distinct adjacent nodes ($\mathcal{Q} \subseteq \{\mathcal{U} \times \mathcal{U}\}$ , and $\phi$ is the incident function that maps each edge to the pair of vertices it connects($\phi:\mathcal{Q} \rightarrow \mathcal{U} \times \mathcal{U}$). 
Given the binary nature $\mathbf{C}$, we propose employing the Hamming distance between the adjacency matrices as the adjacency criterion to structure the topology of $\mathtt{R}$
To this end, we define $\phi$ as
\begin{equation}
\phi : \mathcal{U} \times \mathcal{U} \to \{0,1\}, \quad  
\phi(\mathtt{B},\mathtt{B}^{\prime}) =  
\begin{cases}  
1, & \text{if } d_H(\mathbf{C},\mathbf{C}^{\prime}) \leq m \\  
0, & \text{otherwise.}  
\end{cases}  
\end{equation}
where $\mathbf{C}^{\prime}$ is the $\mathtt{B}^{\prime}$ adjacency matrix, $m$ is the Hamming distance threshold in bits and $d_H(\mathbf{C},\mathbf{C}^{\prime})$ is the Hamming distance function between two distinct adjacency matrices, given by 
\begin{equation}
d_H(\mathbf{C}, \mathbf{C}') = \sum_{k=1}^{K}\sum_{l=1}^{L} \mathbf{1}_{\{c_{k,l} \neq c_{k,l}^{\prime}\}}.
\label{eq:Hdist}
\end{equation}
Consequently, the degree of $\mathtt{B}$ is defined as the number of distinct vertices whose adjacency matrices $\mathbf{C}$ differ from each other by at most $m$ bits, i.e.,
\begin{align} 
    & \deg_m(\mathbf{C}) = \#({\mathbf{C}' \in \mathcal{U} : 1 \leq d_H(\mathbf{C}, \mathbf{C}') \leq m})= \notag \\
    & \deg_m(\mathbf{C}) = \sum_{i=1}^{m} \binom{KL}{i},
\label{eq:dgrC} 
\end{align}
where $\deg(\cdot)$ denotes the degree of a vertex.
This adjacency criterion turns $\mathtt{R}$ into a regular graph, where all vertices have the same degree. Thus, the maximum number of edges $E=\#(\mathcal{E})$ becomes
\begin{equation}   
E = \frac{1}{2}\sum_{u=1}^{U} \deg(\mathtt{B}_u) =\frac{1}{2}\sum_{u=1}^{U} \deg_{m}(\mathbf{C}_u)= \frac{1}{2}{U}\sum_{i=1}^{m} \binom{KL}{i}.
\label{eq:sizeOfE}
\end{equation}

\subsection{Downlink Data Transmission}

Let $s_k \in \mathbb{C}$ denote the data symbol intended for UE $k$, with normalized power $\mathbb{E}\{|s_k|^2\} = 1$, and let $\mathbf{w}_{k,l} = \begin{bmatrix}w_{k,l,1} & \dots & w_{k,l,N} \end{bmatrix}^T \in \mathbb{C}^{N}$ be the normalized precoding vector at AP $l$ for UE $k$, such that $\|\mathbf{w}_{k,l}\|^2 = 1$.
Considering that each AP typically has a few antennas connected to a single power amplifier, 
let $\sqrt{p_{k,l}} \in \mathbb{R}_+$ denote the per-AP power scaling variable to manage the assigned power for UE $k$ at AP $l$, and $\mathbf{P} = [p_{k,l}] \in \mathbb{R}_+^{K \times L}$ denote a given RAN power state. Therefore, the precoded signal transmitted by AP $l$ to UE $k$ is given by $\mathbf{x}_{k,l} = \sqrt{p_{k,l}}\, c_{k,l}\, \mathbf{w}_{k,l} s_k,$ and the total transmit signal from AP $l$ to all UEs is given by $\mathbf{x}_{l} = \sum_{k=1}^K \sqrt{p_{k,l}}\, c_{k,l}\, \mathbf{w}_{k,l} s_k \in \mathbb{C}^{N}$, 
and the total power transmitted by AP $l$ is given by
\begin{equation}
p_{l}(\mathbf{P},\mathbf{C}) = \mathbb{E}\left\{\|\mathbf{x}_l\|^2\right\} 
= \sum_{k=1}^K p_{k,l}c_{k,l} \quad,
\label{eq:totAPpwr}
\end{equation}
restricted by $p_{l}^{\min} \leq p_{l} \leq p_{l}^{\max}$
The transmitted signals propagate over a wireless medium, and the channel vector between the $n$-th antenna of AP $l$ and UE $k$ is denoted by $\mathbf{g}_{k,l} = \begin{bmatrix}g_{k,l,1} & \dots & g_{k,l,N}\end{bmatrix}^T \in \mathbb{C}^{N}$, modeled as $\mathbf{g}_{k,l} = \sqrt{\beta_{k,l}} \cdot \mathbf{h}_{k,l}$ , where $\beta_{k,l} \in \mathbb{R}_+$ is the large-scale fading coefficient (accounting for path loss and shadowing), and $\mathbf{h}_{k,l} \in \mathbb{C}^{N}$ is the small-scale fading vector. Thus, the total received signal at UE $k$ is described by
{\small
\begin{align}
& y_k = 
\sum_{l=1}^L  \mathbf{g}_{k,l}^T \left(\sum_{i=1}^K \sqrt{p_{i,l}}\, c_{i,l}\, \mathbf{w}_{i,l} s_i \right) + n_k = 
\\ \notag
& \underbrace{\sum_{l=1}^L \mathbf{g}_{k,l}^T\, \mathbf{w}_{k,l}\, \sqrt{p_{k,l}}\, c_{k,l} s_k}_{\text{desired signal: } y_{k}^\mathrm{S}} +
\underbrace{\sum_{l=1}^L \mathbf{g}_{k,l}^T \left( \sum_{\substack{i=1 \\ i\neq k}}^K \sqrt{p_{i,l}}\, c_{i,l}\, \mathbf{w}_{i,l} s_i \right)}_{\text{interference: } y_{k}^\mathrm{I}} + n_k,
\end{align}
}where $y_k^{\mathrm{S}}$ and $y_k^{\mathrm{I}}$ are, respectively, the desired signal and interference at UE $k$, and $n_k \sim \mathcal{CN}(0, \sigma^2)$ is the AWGN. The resulting signal-to-interference-plus-noise ratio (SINR) at UE $k$ is 
\begin{equation}
\mathrm{SINR}_k = 
\frac{
\mathbb{E}\left\{ \left| y_k^{\mathrm{S}} \right|^2 \right\}
}{
\mathbb{E}\left\{ \left| y_k^{\mathrm{I}} \right|^2 \right\} + \sigma^2
},
\label{eq:sinrk}
\end{equation}
and the achievable downlink data rate can be expressed as 
\begin{equation}
R_{k}=B\log_{2}{(1+\mathrm{SINR}_{k})}
\label{eq:dwthrp}
\end{equation}
where $B$ is the total downlink bandwidth.

\section{Graph-Based Search and Evaluate Approach}

Many UC CF-mMIMO optimization problems can be formulated as mixed discrete–continuous problems \cite{tds,tds2,rapa}. The presence of binary clustering decisions together with additional discrete and continuous variables typically leads to non-convex formulations, requiring different solution strategies in general-purpose optimizers.
Classical approaches rely on Branch-and-Bound (B\&B) methods combined with relaxation-based techniques such as outer approximation or generalized Benders decomposition \cite{1962:benders:bdecomp}. Although these methods provide optimality guarantees, their repeated branching and relaxation steps result in prohibitive computational cost for large-scale RANs, limiting their applicability in real-time or high-dimensional clustering scenarios.To overcome this, we propose the graph-based search-and-evaluate (GBSE) algorithm, inspired by the Benders decomposition and tailored to a Hamming-based graph topology that exploits RAN domain knowledge (e.g., AP neighborhood relations). 

In this approach, feasible RAN serving states are represented as vertices of a Hamming-based graph. For a given serving state, the remaining continuous and auxiliary discrete subproblems are solved using an appropriate inner solver, yielding the corresponding objective value. The algorithm iteratively transitions to a neighboring serving state that improves the objective function until no improving neighbor exists. This preserves the two-layer decomposition of classical methods while enabling scalable, solver-agnostic optimization without relaxing clustering variables. 

Thus, consider a mixed discrete–continuous optimization problem with objective function $f(\mathbf{C},\mathbf{X},\mathbf{Y})$. Let $\mathbf{C} \in \mathcal{C} \subseteq \{0,1\}^{K\times L}$ denote binary clustering variables defining the RAN serving state, $\mathbf{X} \in \mathcal{X} \subseteq \mathbb{Z}^{K\times L}$ remaining discrete variables, and $\mathbf{Y} \in \mathcal{Y} \subseteq \mathbb{R}^{K\times L}$ continuous variables. Let $g_i(\cdot)$ and $h_j(\cdot)$ denote inequality and equality constraints. The problem can be reformulated through the value function
\begin{equation}
\begin{aligned}
& \Psi(\mathbf{C}) \triangleq
\min_{\mathbf{X} \in \mathcal{X},\text{ }\mathbf{Y} \in \mathcal{Y}}
f(\mathbf{C}, \mathbf{X}, \mathbf{Y}) \\
\text{s.t.} \quad
& g_i(\mathbf{C}, \mathbf{X}, \mathbf{Y}) \leq 0,\;
  h_j(\mathbf{C}, \mathbf{X}, \mathbf{Y}) = 0,
\end{aligned}
\label{eq:modotim}
\end{equation}
where $\Psi(\mathbf{C})$ denotes the optimal value of the inner problem for a fixed serving state. The original problem is therefore $\min_{\mathbf{C} \in \mathcal{C}} \Psi(\mathbf{C})$, where each evaluation of $\Psi(\mathbf{C})$ corresponds to solving the inner problem for a given RAN configuration.

\subsection{Proposed GBSE Algorithm} \label{sec:gbse}

In the proposed GBSE approach, the outer layer is responsible exclusively for the clustering decision, while the inner layer solves the remaining optimization variables associated with a fixed serving state. Accordingly, the outer layer performs two main tasks:
\begin{enumerate*}[label=(\roman*)]
 \item construction of a structured search space and 
 \item efficient exploration of RAN serving states.
\end{enumerate*}
The search space construction starts by retrieving the current RAN serving state $\mathtt{B}^*$, represented by its adjacency matrix $\mathbf{C}^*$, defined as
\begin{multline*}
\begin{aligned}
& \mathbf{C}^* = [c_{k,l}]_{K \times L}, \quad \text{where } 
c_{k,l} =  
\begin{cases}  
1,& \sum_{i=1}^N w_{k,l,i} > 0  \\
0,& \text{otherwise}.
\end{cases}  
\end{aligned}
\end{multline*}
where $\textbf{w}_{k,l}$ is the precoding vector and $w_{k,l,i} \in \textbf{w}_{k,l}$.
The corresponding value function $\Psi(\mathbf{C}^*)$ is then computed by the GBSE inner layer and associated with the current state. The set of neighboring states $\mathbf{C}^\prime$ of $\mathbf{C}^*$ are generated according with the Hamming distance threshold $m$ such as  
\begin{equation}
\mathbf{C}^{\prime} \in \mathcal{N}(\mathbf{C}^*) \iff d_{H}(\mathbf{C}^*,\mathbf{C}^{\prime}) \leq m.
\label{eq:nghdef}
\end{equation}
where $\mathcal{N}(\mathbf{C})$ denotes the set of neighbors of $\mathbf{C}^*$, $d_{H}(\cdot)$ denotes the Hamming distance function given by (\ref{eq:Hdist}).

However, not all Hamming neighbors correspond to feasible RAN configurations. To avoid evaluating forbidden or impractical UE–AP associations, domain knowledge about AP neighborhoods is incorporated. For each UE $k$, the set of candidate APs is restricted to those currently serving the UE and their registered neighboring APs. This information is encoded in a binary mask matrix $\mathbf{M} \in \{0,1\}^{K\times L}$, where entries equal to one indicate positions eligible for flipping. Moreover, 
elementary matrices $\mathbf{E}^{(k,l)}$ based on $\mathbf{M}$ are generated to support logical XOR operation with $\mathbf{C}^*$ in order to produce admissible neighbors as 
\begin{equation}
\mathbf{C}' = \mathbf{C}^* \oplus \mathbf{E}^{(k,l)}, \quad \forall (k,l)\ \text{s.t.}\ \mathbf{M}(k,l)=1,    
\end{equation}
ensuring that only topology-consistent serving states are explored, thereby reducing the search space.

Once the search space is defined, the GBSE outer layer performs the search process. Exhaustive strategies such as BFS or DFS are computationally prohibitive, while informed methods such as A* lack admissible heuristics for nonconvex or fractional objectives. In contrast, Hill-Climbing methods offer a favorable balance between exploration and efficiency. In particular, the Steepest-Ascent Hill-Climbing (SAHC) variant is adopted due to its deterministic updates and predictable complexity \cite{2016:russell:aimodernapproach,2023:Zhang:hcimprove}. At each iteration, all admissible neighbors $\mathbf{C}'$ are evaluated using the inner-layer solver, and the state yielding the best value $\Psi(\mathbf{C}')$ replaces the current state. The process repeats until no improving neighbor exists.

By combining Hamming-based graph modeling with SAHC, the proposed GBSE approach provides
\begin{enumerate*}[label=(\roman*)]
\item an \textbf{outer layer} offering a structured and scalable mechanism for clustering optimization, where the Hamming threshold $m$ controls complexity and convergence, and
\item an \textbf{inner layer} that can adopt any optimization method suited to the utility function and its decision variables (discrete or continuous).
\end{enumerate*}

\subsection{GBSE Algorithm Complexity}

The computational behavior of GBSE over the graph of RAN serving states is characterized by two metrics: the per-iteration runtime and the accumulated runtime until convergence. The per-iteration time accounts for the generation of adjacent serving states and the evaluation of the objective function, and is given by  
$$
t_{iter} = (t_{adjc} + t_{of}) N_{eval};
\quad
t_{acum} = t_{iter}N_{mov},
$$
where $t_{adjc}$ denotes the time required to construct each adjacent state $\mathbf{C}'$ from the current state $\mathbf{C}$, $t_{of}$ is the time required to solve the objective function for a given serving state, $N_{\text{eval}}$ is the number of objective function evaluations per iteration, and $N_{\text{mov}}$ is the number of state transitions until convergence. The overall computational cost is dictated by the topology of the serving-state graph, which regulates the number of neighbors per vertex and, thus, the number of evaluations per iteration.  Hence, for a maximum distance threshold $m$, all vertices share the same degree $\deg_m(\mathbf{C})$, yielding $N_{\text{eval}} = \deg_m(\mathbf{C})$. The per-iteration complexity therefore satisfies  
\begin{equation}
\mathcal{O}(t_{iter})  
= \mathcal{O}((t_{adjc} + t_{of}) \deg_m(\mathbf{C})).  
\end{equation}
The time required to generate an adjacent serving-state matrix $\mathbf{C}'$ scales with the dimension of the binary matrix, yielding $\mathcal{O}(t_{\text{adjc}}) = \mathcal{O}(KL) = \mathcal{O}(n)$, where $n = KL$. The complexity of $t_{\text{of}}$ depends on the inner layer algorithm used to solve the continuous optimization problem. Since an asymptotic analysis focuses on worst-case behavior, we assume that $t_{\text{of}}$ is linear and the per-iteration complexity can be approximated by  
\begin{equation}
\mathcal{O}(t_{iter}) = \mathcal{O}(t_{of} \deg_m(\mathbf{C})).    
\end{equation}
Under a worst-case upper bound in which convergence occurs after visiting all $U$ serving states, the number of state transitions satisfies  $N_{mov} = (U - 1)/\deg_m(\mathbf{C})$. Accordingly, the accumulated runtime is bounded by  
\begin{equation}
\begin{aligned}
& \mathcal{O}(t_{acum})= 
\mathcal{O}(t_{of}\deg_m(\mathbf{C}) \cdot \frac{U - 1}{\deg_m(\mathbf{C})})  
= \mathcal{O}(t_{of}U).  
\end{aligned}
\end{equation}
This result shows that all Hamming-based topologies share the same worst-case accumulated complexity order while differing in per-iteration cost and practical convergence behavior. In the full cell-free case $(D=L)$, the number of serving states scales as $U = 2^{n}$, yielding a worst-case bound of $\mathcal{O}(t_{of}2^{n})$. Restricting the neighborhood size $\deg_m(\mathbf{C})$ enables exploration of local optima with substantially reduced per-iteration complexity.

Other graph topologies were also considered, including linear graphs $(1 \leq \deg(\mathbf{C}) \leq 2)$ and complete graphs $(\deg(\mathbf{C}) = U - 1)$. Linear graphs require low complexity but are highly susceptible to convergence to local optima. Complete graphs correspond to an exhaustive search in a single iteration, equivalent to a Hamming graph with $m = KL$, and are computationally prohibitive for large-scale networks. In contrast, the Hamming-based graph enables a controlled trade-off between complexity and solution quality through appropriate selection of the distance threshold $m$.

\section{GBSE for RAN Energy Efficiency Problem}

In this section, we carry out an energy efficiency (EE) optimization with the proposed framework and GBSE algorithm as an example. The EE optimization quantifies the trade-off between a network utility metric (e.g., SINR or achievable rate) and the total power required to sustain communication and signal processing operations \cite{2015:zappone:eefractional}. We focus on the RAN energy efficiency (RANEE), adopting the aggregate achievable downlink data rate as the utility function. Using (\ref{eq:totAPpwr}) and (\ref{eq:dwthrp}), RANEE is defined as
\begin{equation}
\mathrm{RANEE}(\mathbf{P},\mathbf{C}) = 
\frac{\sum_{k=1}^K R_k(\mathbf{P},\mathbf{C})}{LP_{\mathrm{circuit}}+\sum_{l=1}^L p_{l}(\mathbf{P},\mathbf{C})}.
\label{eq:ree}  
\end{equation}
The numerator is measured in bits per second, while the denominator represents the total power consumption in watts, consisting of the aggregate transmit power $p_l$ across all APs and a constant circuit power term $P_{\mathrm{circuit}}$, assumed identical for all APs. This circuit term accounts for static power consumption associated with baseband processing, fronthaul communication, synchronization, and cooling.

In this context, quality-of-service (QoS) must be guaranteed while minimizing total power expenditure.
Hence, each AP selects its transmit power within prescribed limits $(p_l^{\min}, p_l^{\max})$, and each UE must satisfy a minimum rate requirement $R_{\min}$. The resulting optimization problem jointly determines the transmit powers $p_{k,l}$ and cooperation variables $c_{k,l}$ that maximize RANEE under power and rate constraints. This formulation leads to a mixed-integer non-linear problem (MINLP), characterized by discrete clustering decisions, fractional SINR expressions, and a nonconvex rate–power objective. 
Formally, the RANEE maximization problem is equivalently reformulated as the minimization of the negative RANEE objective such as

\vspace{-1em}
{\small
\begin{equation}
\begin{aligned}
& \min_{\substack{\mathbf{C}\in\{0,1\}^{K\times L} \\ \mathbf{P}\ge0}}
 - \text{RANEE}(\mathbf{P},\mathbf{C})
= - \frac{\sum_{k=1}^{K} R_k(\mathbf{P},\mathbf{C})}
{L P_{\mathrm{circuit}}
+\sum_{l=1}^{L}\sum_{k=1}^{K} c_{k,l} p_{k,l}}
\\
& \text{s.t.} \quad R_k(\mathbf{P},\mathbf{C})\ge R_{\min},\ \forall k,\quad p^{\min}_l\le \sum_{k=1}^{K} c_{k,l}p_{k,l}\le p^{\max}_l,\ \forall l.
\end{aligned}
\label{eq:maxRANEE}
\end{equation} 
}

\subsection{GBSE Application}

The maximization problem proposed in (\ref{eq:maxRANEE}) is a good candidate for applying the GBSE approach due to the discrete decision variable responsible for the UC clustering. The proposed approach first retrieves the current adjacency $\mathbf{C}^*$ and the continuous variable $\mathbf{P}^*$ and maximizes $\text{RANEE}(\mathbf{P}^*,\mathbf{C}^*)$ according to the constraints. In this case, we employ Dinkelbach’s method \cite{2015:zappone:eefractional} to solve the fractional sub-problem in the inner layer, transforming the ratio objective into a sequence of tractable subproblems with subtractive objectives such as
\begin{equation}
\begin{aligned}
& \min_{p \in \mathcal{P}} \quad-[R_{k}(P,C)-\lambda({LP_{\text{circuit}}+\sum_{k=1}^K\sum_{l=1}^L p_{{k,l}}})]
\text{ until} \\
&\lambda^{(t+1)} = \frac{R_{k}(P,C)^t}{({LP_{\text{circuit}}+\sum_{k=1}^K\sum_{l=1}^L p_{{k,l}}})^t} \leq \epsilon_{dinkel}.
\end{aligned}
\end{equation}
where $\lambda^{(t+1)}$ is the next improvement  until it falls below a tolerance $\epsilon_{{dinkel}}$. As soon as the $\Psi(\mathbf{C}^*)\triangleq \text{RANEE}(\mathbf{P}^*,\mathbf{C}^*)$ is retrieved, as in \eqref{eq:modotim}, the outer layer starts to build the $\mathcal{N}(\mathbf{C})$ neighborhood and evaluate the objective function for each element of this set according to  Section \ref{sec:gbse}, leading to 
\begin{equation}
\mathbf{C}^*=\arg\min_{\mathbf{C^{\prime}}\in\mathcal{C}}\quad -\text{RANEE}(\mathbf{P}^*(\mathbf{C^{\prime}}),\mathbf{C^{\prime}}).
\end{equation}
Table \ref{tab:gbsednk} shows the complexity of GBSE while using Dinkelbach's method with some illustrative Hamming distance thresholds $m$. For this comparison, we consider the algorithm complexity of Dinkelbach's method as $\mathcal{O}(n^2)$ \cite{1976:schaible:dnklcomplex}.

\begin{table}[htb]
\centering
\caption{Complexity Analysis of GBSE using Dinkelbach}
\begin{tabular}{|c|c|c|c|}
\hline
\textbf{Graph Topology}           & $deg_m(\mathbf{C})$ & $\mathcal{O}(t_{iter})$   & $\mathcal{O}(t_{acum})$   \\ \hline
Hamming based m=1  & $n$                 & $\mathcal{O}(n^{3})$      & $\mathcal{O}(n^{3}2^{n})$ \\ \hline
Hamming based m=2  & $(n^2+n)/2$         & $\mathcal{O}(n^{4})$      & $\mathcal{O}(n^{3}2^{n})$ \\ \hline
Hamming based m=3  & $(n^3+5n)/6$         & $\mathcal{O}(n^{5})$      & $\mathcal{O}(n^{3}2^{n})$ \\ \hline
Hamming based m=KL & $2^n-1$             & $\mathcal{O}(n^{3}2^{n})$ & $\mathcal{O}(n^{3}2^{n})$ \\ \hline
\end{tabular}
\label{tab:gbsednk}
\end{table}

\subsection{Numerical Evaluation}

We evaluate the proposed framework and the GBSE algorithm combined with Dinkelbach’s method using RAN energy efficiency (RANEE) as the objective. GBSE with Hamming distance thresholds $m \in \{1,2,3\}$ is compared against three baselines: joint optimization via relaxation-and-rounding (JO), exhaustive search (ExSrch), and a channel-norm-based (chNm) criterion \cite{ucentric} evaluated under the same Hamming distance thresholds. Both GBSE and ExSrch operate within the proposed framework, whereas chNm replaces RANEE with the channel norm as the objective function. Performance is assessed in terms of elapsed processing time and the optimality gap in RANEE relative to ExSrch.

\begin{figure}
\centering
\begin{tikzpicture}
\begin{groupplot}[
    group style={
        group size=2 by 1,      
        horizontal sep=1cm,     
        x descriptions at=edge bottom
    },
    width=5cm,
    height=5cm,
    xlabel style={font=\scriptsize},
    ylabel style={font=\scriptsize},
    xticklabel style={font=\scriptsize},
    yticklabel style={font=\scriptsize},
]

\nextgroupplot[
    ylabel={Avg Normalized RANEE [bit/J]},
    xlabel={Density (K/(L*N))},
    xmin=0.01,xmax=2.6,
    ymin=0.65, ymax=1.1,
    grid=both,
]

\addplot [dashed,ultra thick,color=orange] coordinates {(0.041666667,1)(0.05,1)(0.0625,1)(0.083333333,1)(0.1,1)(0.125,1)(0.15,1)(0.166666667,1)(0.1875,1)(0.2,1)(0.208333333,1)(0.25,1)(0.3,1)(0.3125,1)(0.333333333,1)(0.375,1)(0.4,1)(0.416666667,1)(0.5,1)(0.625,1)(0.666666667,1)(0.75,1)(0.833333333,1)(1,1)(1.25,1)(1.5,1)(2,1)(2.5,1)};

\addplot [color=red] coordinates {(0.041666667,1)(0.05,0.988867476)(0.0625,0.975242368)(0.083333333,0.957676548)(0.1,0.946544024)(0.125,0.932918916)(0.15,0.921786392)(0.166666667,0.915353096)(0.1875,0.908161284)(0.2,0.904220572)(0.208333333,0.901727988)(0.25,0.890595464)(0.3,0.87946294)(0.3125,0.876970356)(0.333333333,0.873029644)(0.375,0.865837832)(0.4,0.86189712)(0.416666667,0.859404536)(0.5,0.848272012)(0.625,0.834646904)(0.666666667,0.830706192)(0.75,0.82351438)(0.833333333,0.817081084)(1,0.80594856)(1.25,0.792323452)(1.5,0.781190928)(2,0.763625108)(2.5,0.75)};

\addplot [color=blue] coordinates {(0.041666667,1)(0.05,0.987531573)(0.0625,0.972271452)(0.083333333,0.952597734)(0.1,0.940129307)(0.125,0.924869186)(0.15,0.912400759)(0.166666667,0.905195468)(0.1875,0.897140638)(0.2,0.892727041)(0.208333333,0.889935347)(0.25,0.87746692)(0.3,0.864998493)(0.3125,0.862206798)(0.333333333,0.857793202)(0.375,0.849738371)(0.4,0.845324775)(0.416666667,0.84253308)(0.5,0.830064653)(0.625,0.814804532)(0.666666667,0.810390935)(0.75,0.802336105)(0.833333333,0.795130814)(1,0.782662387)(1.25,0.767402266)(1.5,0.754933839)(2,0.735260121)(2.5,0.72)};

\addplot [dashed,color=blue] coordinates {(0.041666667,1)(0.05,0.991539282)(0.0625,0.981184199)(0.083333333,0.967834177)(0.1,0.959373458)(0.125,0.949018376)(0.15,0.940557658)(0.166666667,0.935668353)(0.1875,0.930202576)(0.2,0.927207635)(0.208333333,0.925313271)(0.25,0.916852553)(0.3,0.908391834)(0.3125,0.90649747)(0.333333333,0.90350253)(0.375,0.898036752)(0.4,0.895041811)(0.416666667,0.893147447)(0.5,0.884686729)(0.625,0.874331647)(0.666666667,0.871336706)(0.75,0.865870929)(0.833333333,0.860981624)(1,0.852520906)(1.25,0.842165823)(1.5,0.833705105)(2,0.820355082)(2.5,0.81)};

\addplot [mark=o,color=blue] coordinates {(0.041666667,1)(0.05,0.994211088)(0.0625,0.987126031)(0.083333333,0.977991805)(0.1,0.972202893)(0.125,0.965117836)(0.15,0.959328924)(0.166666667,0.95598361)(0.1875,0.952243867)(0.2,0.950194698)(0.208333333,0.948898554)(0.25,0.943109641)(0.3,0.937320729)(0.3125,0.936024585)(0.333333333,0.933975415)(0.375,0.930235672)(0.4,0.928186503)(0.416666667,0.926890359)(0.5,0.921101446)(0.625,0.91401639)(0.666666667,0.91196722)(0.75,0.908227477)(0.833333333,0.904882164)(1,0.899093251)(1.25,0.892008195)(1.5,0.886219282)(2,0.877085056)(2.5,0.87)};

\addplot [color=green] coordinates {(0.041666667,1)(0.05,0.98619567)(0.0625,0.969300536)(0.083333333,0.94751892)(0.1,0.93371459)(0.125,0.916819456)(0.15,0.903015126)(0.166666667,0.895037839)(0.1875,0.886119992)(0.2,0.881233509)(0.208333333,0.878142705)(0.25,0.864338375)(0.3,0.850534045)(0.3125,0.847443241)(0.333333333,0.842556759)(0.375,0.833638911)(0.4,0.828752429)(0.416666667,0.825661625)(0.5,0.811857295)(0.625,0.794962161)(0.666666667,0.790075679)(0.75,0.781157831)(0.833333333,0.773180544)(1,0.759376215)(1.25,0.74248108)(1.5,0.728676751)(2,0.706895134)(2.5,0.69)};

\addplot [dashed,color=green] coordinates {(0.041666667,1)(0.05,0.987086272)(0.0625,0.971281147)(0.083333333,0.950904796)(0.1,0.937991068)(0.125,0.922185942)(0.15,0.909272215)(0.166666667,0.901809592)(0.1875,0.893467089)(0.2,0.888895864)(0.208333333,0.886004466)(0.25,0.873090738)(0.3,0.86017701)(0.3125,0.857285613)(0.333333333,0.852714387)(0.375,0.844371885)(0.4,0.83980066)(0.416666667,0.836909262)(0.5,0.823995534)(0.625,0.808190408)(0.666666667,0.803619183)(0.75,0.795276681)(0.833333333,0.787814058)(1,0.77490033)(1.25,0.759095204)(1.5,0.746181476)(2,0.725805126)(2.5,0.71)};

\addplot [mark=o,color=green] coordinates {(0.041666667,1)(0.05,0.988422175)(0.0625,0.974252062)(0.083333333,0.95598361)(0.1,0.944405785)(0.125,0.930235672)(0.15,0.918657847)(0.166666667,0.91196722)(0.1875,0.904487735)(0.2,0.900389395)(0.208333333,0.897797107)(0.25,0.886219282)(0.3,0.874641458)(0.3125,0.87204917)(0.333333333,0.86795083)(0.375,0.860471345)(0.4,0.856373005)(0.416666667,0.853780718)(0.5,0.842202893)(0.625,0.82803278)(0.666666667,0.82393444)(0.75,0.816454955)(0.833333333,0.809764328)(1,0.798186503)(1.25,0.78401639)(1.5,0.772438565)(2,0.754170113)(2.5,0.74)};

\nextgroupplot[
    ylabel={Avg Processing Time [s]},
    xlabel={NE (K+L+N)},
    ymin=3,ymax=50,
    grid=both,
    legend style={at={(-0.25,-0.25)},anchor=north,legend columns=4,font=\scriptsize},
]

\addplot [dashed,ultra thick,color=orange,restrict y to domain=3:150] coordinates {(4,5)(5,11.3981637)(6,25.98362715)(7,59.23312715)(8,135.029776)(9,307.8182981)(10,701.7126704)(11,1599.647178)(12,3646.608079)(13,8312.927166)(14,18950.42093)(15,43200)};
\addlegendentry{ExSrch}

\addplot [color=red] coordinates {(4,3.2)(5,4.17712399)(6,5.452614009)(7,7.117576496)(8,9.290937355)(9,12.12793666)(10,15.83121724)(11,20.66529917)(12,26.97547404)(13,35.21246867)(14,45.96463989)(15,60)};
\addlegendentry{JO}

\addplot [color=blue] coordinates {(4,3.3)(5,4.315702999)(6,5.644027992)(7,7.381196524)(8,9.653046053)(9,12.62414539)(10,16.5097158)(11,21.5912212)(12,28.236757)(13,36.9277142)(14,48.29365058)(15,63.15789474)};
\addlegendentry{GBSE-1}

\addplot [dashed,color=blue] coordinates {(4,3.5)(5,4.599110964)(6,6.043377616)(7,7.941189786)(8,10.43497515)(9,13.71188817)(10,18.01785578)(11,23.67603373)(12,31.11105894)(13,40.88091779)(14,53.71882207)(15,70.58823529)};
\addlegendentry{GBSE-2}

\addplot [mark=o,color=blue] coordinates {(4,4)(5,5.239579129)(6,6.863297361)(7,8.990197402)(8,11.77621267)(9,15.42559953)(10,20.20591233)(11,26.46761913)(12,34.6697962)(13,45.41378514)(14,59.48728019)(15,77.92207792)};
\addlegendentry{GBSE-3}

\addplot [color=green] coordinates {(4,3.2)(5,4.192654475)(6,5.493234858)(7,7.197261158)(8,9.429884125)(9,12.35507684)(10,16.18767756)(11,21.20916836)(12,27.78834832)(13,36.40841966)(14,47.70247613)(15,62.5)};
\addlegendentry{chNm-1}

\addplot [dashed,color=green] coordinates {(4,3.5)(5,4.584631738)(6,6.005385191)(7,7.866422727)(8,10.30418608)(9,13.49739959)(10,17.68017329)(11,23.15916674)(12,30.33607167)(13,39.73706199)(14,52.05137016)(15,68.18181818)};
\addlegendentry{chNm-2}

\addplot [mark=o,color=green] coordinates {(4,3.8)(5,4.989220966)(6,6.550612064)(7,8.600645013)(8,11.29224169)(9,14.82618131)(10,19.46607754)(11,25.55804267)(12,33.55650588)(13,44.05811123)(14,57.84622428)(15,75.94936709)};
\addlegendentry{chNm-3}
\end{groupplot}
\end{tikzpicture}
\caption{Evaluation of RANEE and Processing Time}
\label{fig:plotSNWwclfinal}
\end{figure}
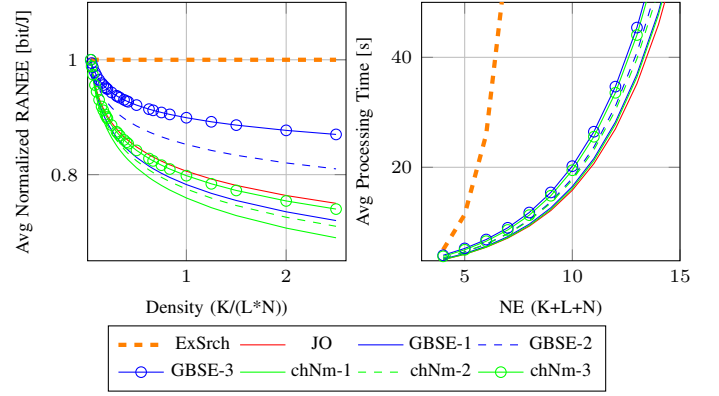

Due to its prohibitive complexity, ExSrch is evaluated only for a small UC CF-mMIMO scenario with $L=6$ APs serving up to $K=5$ UEs. APs are uniformly deployed over a $200 \times 200$ m$^2$ dense urban area. AP neighborhoods are defined via a received power threshold. AP locations are fixed, while UE positions are randomly generated. For each realization, both the UE deployment seed and the AP antenna configuration (2 or 4 antennas) are varied. Table \ref{tbl:simparam} shows the parameters used in all evaluations and results are reported in Fig.~\ref{fig:plotSNWwclfinal}.

As expected, ExSrch consistently achieves the global optimal RANEE but exhibits exponentially increasing processing time, making it impractical beyond small networks. JO achieves significantly lower runtimes but suffers from increasing performance degradation as network size grows. {The chNm criterion yields lower RANEE than JO, even for $m=3$, while exhibiting comparable runtimes}. GBSE offers a clear trade-off between complexity and optimality. For $m=1$, GBSE achieves runtimes comparable to JO, at the expense of a larger optimality gap. Increasing the Hamming distance threshold ($m=2$ and $m=3$) systematically improves the achieved RANEE, approaching the global optimum with moderate increases in processing time. 

\begin{table}[htb]
\caption{Simulation parameters}
\scriptsize
\setlength{\tabcolsep}{4pt}
\renewcommand{\arraystretch}{0.95}

\begin{subtable}[t]{0.45\linewidth}
\centering
\caption{AP Parameters}
\begin{tabular}{lc}
\hline
height ($h_l$) & 6 m \\
carrier frequency ($f$) & 2 GHz \\
carrier bandwidth ($BW$) & 10 MHz \\
precoding scheme & MRT \\
min TX power ($p_l^{\min}$) & 0.01 W \\
max TX power ($p_l^{\max}$) & 0.2 W \\
circuit power ($P_{\text{circuit}}$) & 0.05 W \\
Noise figure ($N_l$) & 9 dB \\
\hline
\end{tabular}
\end{subtable}
\begin{minipage}[t]{0.47\linewidth}

\begin{subtable}[t]{\linewidth}
\centering
\caption{UE Parameters}
\begin{tabular}{lc}
\hline
height ($h_k$) & 1.5 m \\
RX sensitivity & $-100$ dBm \\
min DL throughput ($R_{\min}$) & 1 Mbps \\
\hline
\end{tabular}
\end{subtable}

\vspace{6pt}

\begin{subtable}[t]{\linewidth}
\centering
\caption{Link Budget}
\begin{tabular}{lc}
\hline
propagation model & UMi (3GPP 38.901) \\
target SNR & $-5$ dB \\
shadowing std dev & 6 dB \\
\hline
\end{tabular}
\end{subtable}

\end{minipage}
\label{tbl:simparam}
\end{table}

\section{Conclusion}

We introduced a framework for mixed integer–continuous optimization in UC CF-mMIMO networks based on a structured RAN serving-state graph. By separating discrete clustering decisions from  the remaining resource allocation variables, the framework allows integration with general-purpose solvers. Hamming-based topologies ensure controllable complexity and scalability. A GBSE algorithm was developed to operate on this structure. An energy-efficiency maximization example using GBSE demonstrated improved performance over previous approaches, and the graph formulation is expected to further benefit from parallel and distributed processing.

\bibliographystyle{IEEEtran}
\bibliography{ref}

\end{document}